\documentclass{WileyMSP-template}
\usepackage{amsmath,amssymb,graphicx,hyperref,float}
\usepackage{xcolor}
\usepackage{hyperref}
\usepackage{quantikz}
\usepackage{subcaption}
\hypersetup{
colorlinks,
linkcolor={red!50!black},
citecolor={blue!80!black},
urlcolor={blue!80!black}
}

\usepackage[percent]{overpic}

\begin{document}

\pagestyle{fancy}
\rhead{\includegraphics[width=2.5cm]{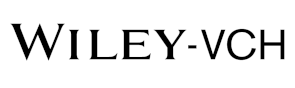}}

\title{Designing Majorana Quasiparticles in InAsP Quantum Dots in InP Nanowires with  Variational Quantum Eigenvalue Solver}

\maketitle
% Author: Please give full first and last names for authors and include * after the name of all corresponding authors

\author{Mahan Mohseni*},
\author{Iann Cunha*},
\author{Daniel Miravet*},
\author{Alina Wania Rodrigues*},
\author{Hassan Allami*},
\author{Ibsal Assi*},
\author{Marek Korkusinski*},
%\author{Stefanie Czischek*},
\author{and Pawel Hawrylak*}

% Dedication
\dedication{}

% Affiliations: Please provide adacemic titles (Prof. or Dr.) for all authors where applicable, and include an institutional email address for all corresponding authors
\begin{affiliations}
Mahan Mohseni\\
Department of Physics, University of Ottawa, Ottawa, Canada K1N 6N5\\
mghaf073@uottawa.ca

Dr. Ibsal Assi\\
Department of Physics, University of Ottawa, Ottawa, Canada K1N 6N5\\
iassi@uottawa.ca

Dr. Iann Cunha\\
Department of Physics, University of Ottawa, Ottawa, Canada K1N 6N5\\
icunha@uottawa.ca

Dr. Daniel Miravet\\
Department of Physics, University of Ottawa, Ottawa, Canada K1N 6N5\\
dmiravet@uottawa.com

Alina Wania Rodrigues\\
Department of Physics, University of Ottawa, Ottawa, Canada K1N 6N5\\
awaniaro@uottawa.ca

Dr. Hassan Allami\\
Department of Physics, University of Ottawa, Ottawa, Canada K1N 6N5\\
mallami@uottawa.ca

Prof. Marek Korkusinski\\
Quantum and Nanotechnology Research Center, National Research Council, Ottawa, Canada K1A0R6\\
marek.korkusinski@nrc-cnrc.gc.ca

%Prof. Stefanie Czischek\\
%Department of Physics, University of Ottawa, Ottawa, Canada K1N 6N5\\
%stefanie.czischek@uottawa.ca

Prof. Pawel Hawrylak\\
Department of Physics, University of Ottawa, Ottawa, Canada K1N 6N5\\
pawel.hawrylak@uottawa.ca
\end{affiliations}

% Keywords: Please provide a minimum of three and a maximum of seven keywords, separated by commas


% Abstract should be written in the present tense and impersonal style (i.e., avoid we), and be at most 200 words long
\begin{abstract}
This work presents steps toward the design of  Majorana zero modes (MZM) in InAsP quantum dots (QD) embedded in an InP semiconducting nanowire in contact with a p-type superconductor described by the Kitaev Hamiltonian. The single particle spectrum is obtained from million atom atomistic calculations with QNANO and many-electron spectra using exact diagonalization (ED) and the hybrid Variational Quantum Eigensolver (VQE) method. A variational ansatz is constructed to capture the ground state of the system by utilizing a generalized form of the analytical solution for a particular set of parameters. By systematically deviating from the analytically solvable regime while maintaining the system in the topological phase (TP), the effectiveness of the variational function in reproducing the correct ground state and topological properties of the system is evaluated. This is done through a quantum algorithm for a many-body state containing MZM. The results are compared with exact solution in topological phase and demonstrate the capability of VQE, along with classical simulations, to accurately model the many-body spectra in topologically nontrivial state. 
\end{abstract}

% Text: Please use section headings and subheadings as specified below. For communications, all section headings apart from Experimental Section should be removed
% Please make the first reference to a display item bold: \textbf{Figure 1}
% Do not abbreviate Figure, Equation, etc.; display items are always singular, i.e., Figure 1 and 2.
% Equations are always singular, i.e., Equation 1 and 2, and should be inserted using the {equation} environment, not as graphics
% Please do not use footnotes in the text, additional information can be added to the Reference list.

\section{Introduction}
We explore here an array of InAsP semiconductor QDs in InP nanowire as a host for Majorana quasiparticles for topological quantum computation. Topological quantum computation \cite{freedman2003topological, das2006topological, marra2022majorana, sarma2015majorana,field2018introduction}
is a robust framework for realizing fault-tolerant quantum computing \cite{gottesman1998theory, preskill1998fault, shor1996fault, kitaev2003fault}, which aims to perform computations reliably despite the presence of noise and errors \cite{marra2022majorana, braunstein2001scalable, cirac2000scalable}.

A promising candidate for implementing this protection is MZM \cite{kim2012majorana}. MZMs are quasiparticles that arise in topological phases of matter, particularly in systems that combine strong spin-orbit coupling with superconductivity \cite{flensberg2021engineered}. Their non-abelian statistics enable fault-tolerant quantum gatesm through the process of braiding-exchanging the positions of MZM modes. Unlike conventional qubits, which are vulnerable to local noises, the information stored in MZMs is distributed non-locally, making it protected from decoherence. 

A candidate to host MZMs is the 1D chain of InAs QDs in InP nanowire \cite{koong2020multiplexed,laferriere2021systematic} containing spinless fermions with nearest-neighbour hopping and p-wave superconducting pairing, namely the Kitaev chain \cite{kitaev2003fault, kitaev2001unpaired, mohseni2023majorana}. In its topological phase (TP), the Kitaev chain hosts MZMs at its ends, where the fermionic excitations split into two spatially separated Majorana particles, each localized at opposite ends of the chain \cite{kitaev2003fault, kitaev2001unpaired, mohseni2023majorana}. 

We describe here one electron properties of the nanowire using QNANO code \cite{ zielinski2010atomistic, cygorek2020atomistic} and many-body properties using ED on a classical computer \cite{mohseni2023majorana} and VQE on a quantum computer.

While ED method provides exact solutions, its applicability is limited by the exponential growth of the Hilbert space. For instance, for chain with $N=25$ QDs the Hamiltonian matrix has the dimension  $2^{25}$. Hence ED is limited to relatively small systems. In this study, we try to find an estimate of the ground state energy of the Kitaev Hamiltonian on a quantum computer \cite{thakurathi2014majorana, chen2014majorana}, using the VQE method. We use analytical and numerical ED to compute the full energy spectrum \cite{mohseni2023majorana} and then implement VQE to compute even and odd ground states by constructing a quantum circuit guided by the analytical solution \cite{mohseni2023majorana}. By comparing the VQE results with those from ED, we demonstrate that VQE provides an accurate estimate of the ground state energy for large systems.

\section{Semiconductor Nanowire as Kitaev Chain}
\begin{figure}[htbp]
\centering
  \includegraphics[width=\textwidth]{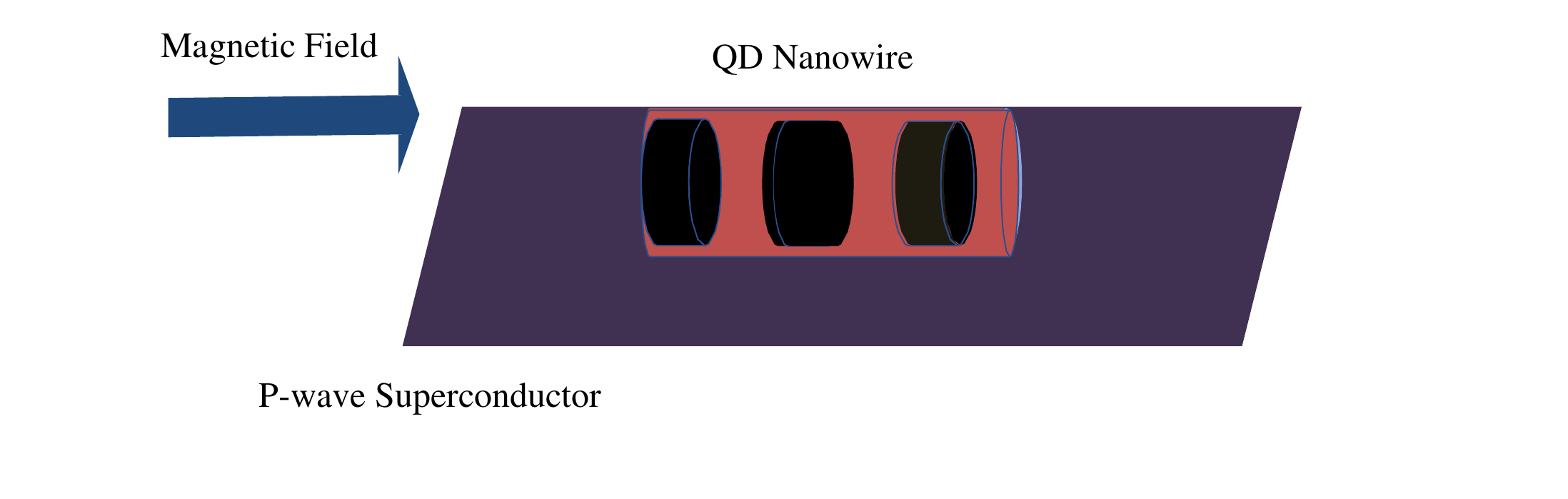}
  \caption{Schematic visualization of the chain of QDs (in black) in a nanowire (red). The QD nanowire is placed on top of a p-wave superconductor (violet). The external magnetic field (blue) is along the nanowire.}
  \label{schem_ham}
\end{figure}

\begin{figure}[htbp]
    \centering
    \begin{subfigure}[b]{0.33\textwidth}
        \centering
        \includegraphics[width=\textwidth]{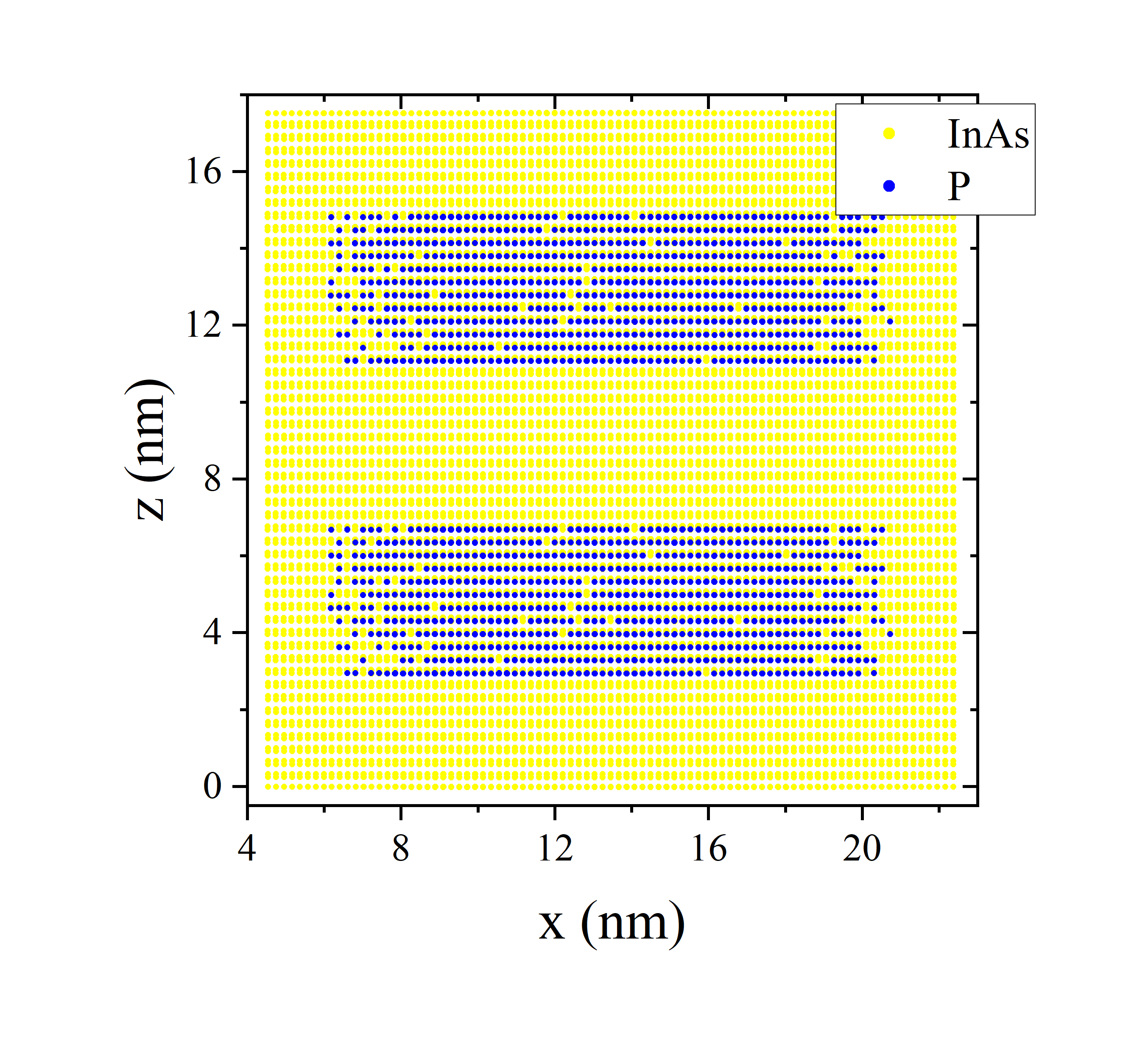}
        \caption{}
        \label{dQDs_a}
    \end{subfigure}
    \begin{subfigure}[b]{0.33\textwidth}
        \centering
        \includegraphics[width=\textwidth]{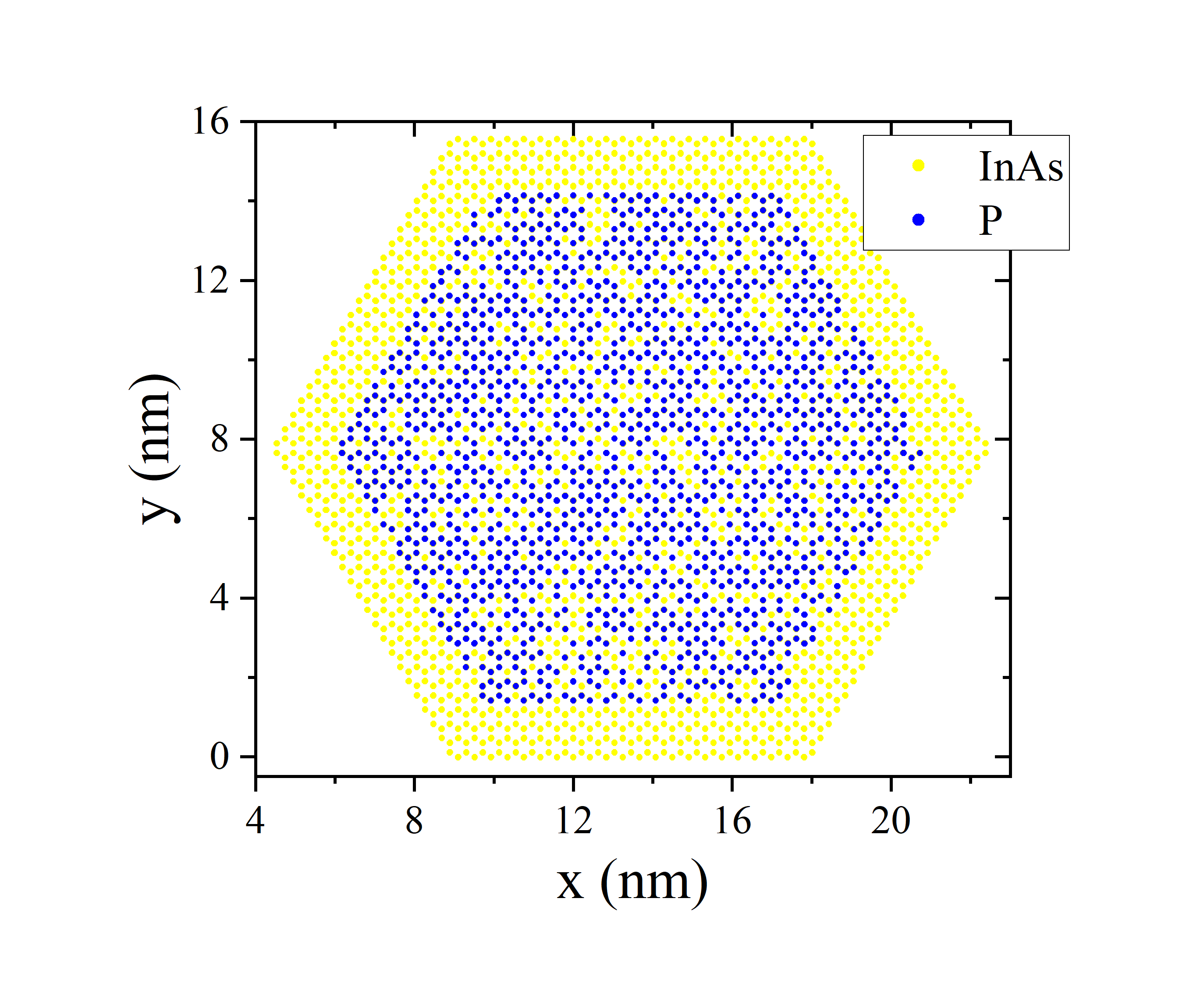}
        \caption{}
        \label{dQDs_b}
    \end{subfigure}
    \begin{subfigure}[t]{0.3\textwidth}
        \centering
        \includegraphics[width=\textwidth]{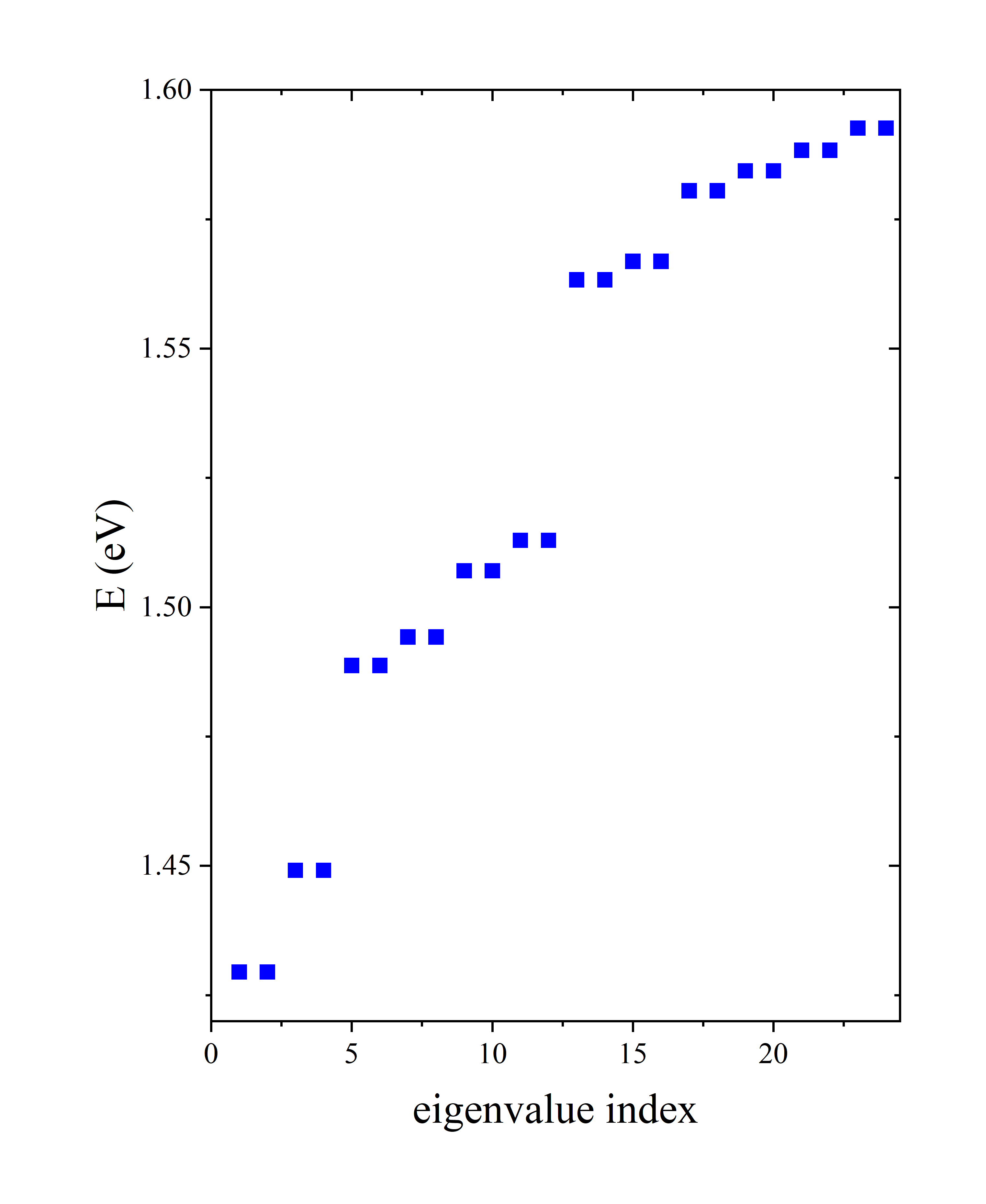}
        \caption{}
        \label{dQDs_c}
    \end{subfigure}
    \caption{(a) Side cross sections of double QDs nanowire. (b) Top cross sections of double QDs nanowire. (c) Single particle spectrum of conduction band electron in a double quantum dot. The splitting of two degenerate s-levels due to tunneling is visible. The degeneracy is due to spin.}
    \label{dQDs}
\end{figure}

The Kitaev chain considered in this work is implemented in a hexagonal InP nanowire with an array of embedded InAsP QDs in the proximity of a p-wave superconductor and in the presence of applied external magnetic field \cite{mohseni2023majorana, talantsev2019p,wang2019theory,yuan2014possible,frigeri2004superconductivity,hardy2005p,ishida1998spin}. The QD nanowire is schematically shown in Figure \ref{schem_ham}.
We performed microscopic calculations of single particle spectrum using QNANO code \cite{ zielinski2010atomistic, cygorek2020atomistic} for electron in one and two InAsP QDs embedded within an InP nanowire to determine the energy of the lowest conduction band states in a QD and interdot tunneling $t$ in the InAsP QD array. QNANO calculations are based on ab initio-based tight-binding model, where the QD nanowire is created by first establishing an InP matrix and defining a hexagonal nanowire within it, as illustrated in Figure \ref{dQDs_a} and \ref{dQDs_b}. Next, random P atoms are replaced with As atoms at a concentration of 20\%. Figure \ref{dQDs_c} displays the energy levels of the electron in the conduction band of a double QD, highlighting the characteristics s, p, and d shells of an isolated QD, which are split due to the tunneling between the two QDs. Despite the random distribution of As atoms, the spectrum reveals a well-defined s shell followed by two states of the p shell \cite{laferriere2021systematic, manalo2024microscopic, manalo2021electronic}.
 
% By applying a strong magnetic field along the nanowire, we can isolate the low energy spin component and treat those electrons as spinless. The Kitaev Hamiltonian for $N$ QDs is the extended Hubbard model \cite{mohseni2023majorana,  manalo2024microscopic, manalo2021electronic, sayyad2023topological}. The Hamiltonian for spinfull electrons in a QD chain in a nanowire placed on top of an s-type superconducting substrate with effective spin-orbit coupling between the dots has been discussed in \cite{stoudenmire2011interaction}. For p-wave superconductor we start with non-interactive Kitaev Hamiltonian \cite{kitaev2001unpaired, mohseni2023majorana}
By applying a strong magnetic field along the nanowire, we can isolate the low energy spin component and treat those electrons as spinless. In this context, for p-wave superconductor the non-interactive Kitaev Hamiltonian \cite{kitaev2001unpaired, mohseni2023majorana} is given as
\begin{equation}
    \hat{H} = t\sum_{i=1}^{N-1} ( c^\dagger_{i+1} c_i +\text{h.c.}  ) + \Delta \sum_{i=1}^{N-1} ( c^\dagger_{i+1}c^\dagger_i +\text{h.c.} )
		-\mu \sum_{i=1}^{N} c^\dagger_{i}c_i , \label{ham}
\end{equation}
where $c_i^\dagger (c_i)$ is the fermionic creation (annihilation) operator of an electron in dot $i$, $t$ is tunneling matrix element between adjacent conduction s-levels, $\Delta$ is the pairing energy for two electrons on adjacent conduction levels, and $\mu$ is the chemical potential measured from the conduction energy level. This Hamiltonian is valid as long as the $t$ and $\Delta$ are small compared with energy levels of s and p-shell. 

\subsection{Exact Diagonalization and Many-body Spectrum of Kitaev Chain}
The many-body spectrum of the Kitaev Hamiltonian was obtained earlier in \cite{mohseni2023majorana} and we summarise it here for completeness. The ED method involves expanding the wave function in terms of all configurations for even and odd number of electrons, constructing the full Hamiltonian matrix in the space of configurations and diagonalizing it to obtain its eigenvalues and eigenvectors, providing exact solutions for finite-sized systems. In case of $N$ spinless orbitals, there are $2^N$ configurations constructed as
\begin{equation}
        \ket{\alpha_1 \dots \alpha_N} = \prod_{i=1}^N (c_i^\dagger)^{\alpha_i} \ket{0} ,
        \label{eq:kitaev_confs}
    \end{equation}
where $\ket{0}$ is the vacuum of electrons, $\alpha_i = 1 \text{ or } 0$, corresponds to having 1 or 0 electron in orbital $i$.

The pairing term in the Kitaev Hamiltonian changes the particle number but conserves parity, hence the eigenstates are linear combinations of electronic configurations with different electron numbers but given parity:
\begin{equation}
        \ket{\psi^\nu} = \sum_{M,p_M} C^\nu_{M,p_M}\ket{{M,p_M}},
      \label{eq:ED_general_psi}
    \end{equation}
where we are populating $N$ sites with $M=0,1,\dots , N$ electrons, and for a given number of electrons $M$ we generate electron configurations $p_M$. In order to obtain the coefficients $C^\nu_{M,p_M}$, we apply the Hamiltonian to this state, and by using the orthogonality of the configurations we obtain the eigenvalue equation
\begin{equation}
      \sum_{p_M,M} \bra{q_{M'},M'} \hat{H} \ket{p_M,M} C^\nu_{M,p_M} = E^\nu C^\nu_{M',q_{M'}}.
     \label{eq:Wfn_MatrixH}
\end{equation}

Since the Kitaev Hamiltonian in Equation (\ref{ham}) only changes particle number in pairs,
the matrix element $\bra{q_{M'},M'} \hat{H} \ket{p_M,M}$ is non-zero only if $M$ and $M'$ have the same parity, i.e., if they are both even or odd. This parity symmetry allows us to break the Hilbert space into two decoupled subspaces of even and odd configurations.

\begin{figure}[H]
\centering
  \includegraphics[width=0.87\linewidth]{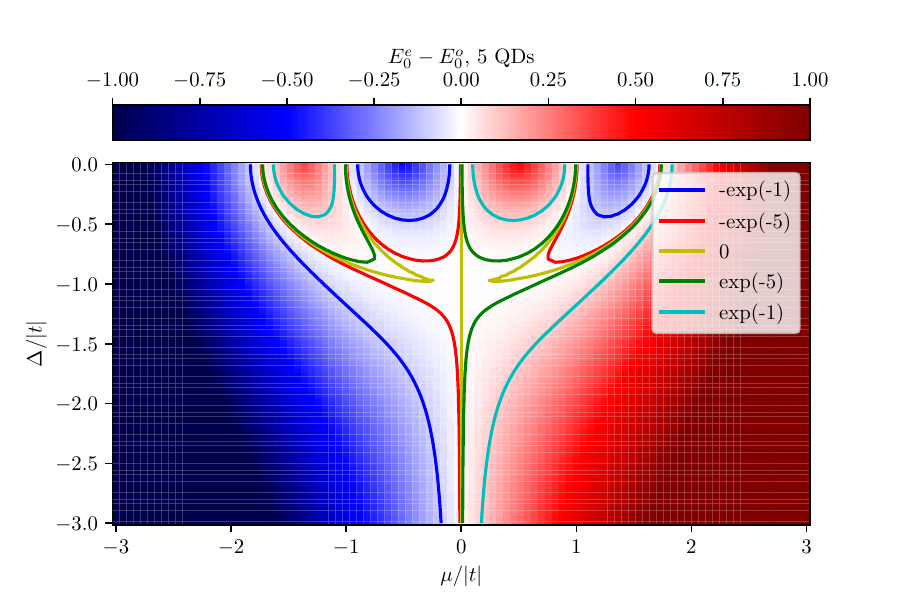}
  \caption{Ground state energy difference of the even and odd subspaces for a chain of 5 QDs, by using ED, with $t=-1$. The contours show the boundaries between the trivial and the TP with various precision, based on the size of the chain.}
  \label{TP}
\end{figure}

Results of the ED for $N=5$ QDs are shown in Figure \ref{TP}, where the energy difference between the even and odd electron numbers is plotted \cite{mohseni2023majorana}. The system is in TP as long as the difference is zero. The MZMs appear when the system is in the TP, where the ground state is doubly degenerate , with one state corresponding to the even and the other one corresponding to the odd subspace. The TP is centered at $t=\Delta$ and $\mu=0$ \cite{sarma2015majorana, mohseni2023majorana}, for which analytical solution exists. It can be shown that as long as $2|t|>\mu$, the system remains in the TP \cite{kitaev2001unpaired, leijnse2012introduction}, as seen in Figure \ref{TP}. The numerical ED can be carried out for small chains for any set of parameters, but alternative methods have to be applied for large systems.
We now turn to the main objective of this work, i.e., the application of VQE method and quantum computer to determining the ground states for even/odd number of electrons
and presence or absence of the TP with MZMs.

\section{Variational Quantum Eigensolver for a Kitaev chain}
The VQE is a hybrid method that aims to estimate the ground state energy of a given Hamiltonian through a quantum-classical computational approach \cite{fedorov2022vqe, tilly2022variational}. The benefit of quantum comes from the capacity to manage and manipulate the variational states as quantum circuits \cite{elfving2020will, gonthier2020identifying}. The variational state is generated by a unitary operation (quantum circuit) acting on an initial state. By using a classical optimizer to fine-tune the variational parameters of the quantum circuit, VQE minimizes the energy for large systems \cite{pellow2021comparison, ravi2022cafqa, wierichs2020avoiding}.

The first step in the VQE workflow is to express the Hamiltonian \(\hat{H}\) in terms of Pauli operators, through the Jordan-Wigner (JW) transformation. A quantum circuit with adjustable gates, parameterized by parameters  \(\boldsymbol{\theta}\), is then applied to generate the quantum state \(\ket{\psi(\boldsymbol{\theta})}\), where \(\boldsymbol{\theta}\) represents a set of parameters used in a quantum circuit, allowing the Hamiltonian written in Pauli operators to act on the state \(\ket{\psi(\boldsymbol{\theta})}\). The expectation value of the energy is measured using the quantum computer as follows:

\begin{equation}
E(\boldsymbol{\theta}) = \bra{\psi(\boldsymbol{\theta})} \hat{H} \ket{\psi(\boldsymbol{\theta})}.
\end{equation}

The measured energy \(E(\boldsymbol{\theta})\) is fed into the classical optimizer, which adjusts the parameters \(\boldsymbol{\theta}\) to minimize the energy. This process is iterated, repeating the quantum circuit execution and the measurement, with the optimizer refining \(\boldsymbol{\theta}\) until the minimum energy is found. Ultimately, the optimized parameters correspond to the approximate ground state of the Hamiltonian, providing a solution to the problem.

\subsection{Jordan-Wigner (JW) Transformation-Kitaev Chain on a Quantum Computer}
The JW transformation maps fermionic operators
into spin operators represented by the Pauli matrices, and spin operators into qubits of a quantum computer  \cite{batista2001generalized, fradkin1989jordan, nielsen2005fermionic}.

The JW transformation for spinless fermions is defined as
\begin{subequations}
\begin{align}
    &c_j^\dagger = e^{-i \Phi_j} S_j^+,\\
    &c_j = S_j^- e^{i \Phi_j},
\end{align}
\end{subequations}
where $S^{\pm}_{j}=S_{j}^{x}\pm iS_{j}^{y}$ are the rising and lowering spin operators for site $j$, and $\Vec{S}=\frac12 \Vec{\sigma}$ are the spin-$\frac12$ operators. The phase factor $\Phi_j$ can be express as 
\begin{equation}
\Phi_j = \pi \sum_{m=1}^{j-1} \left( \frac{1}{2} + S_m^z \right) ,
\end{equation}
which introduces non-locality in the mapping, ensuring that the fermionic operators obey the correct anti-commutation relations. Therefore, by using the exponential term as \(e^{\pm i\Phi_j} = \prod_{m=1}^{j-1} \left(-2S_m^z\right)\), as previously demonstrated in \cite{galitski2010fermionization, al2023prototypical}, the fermionic products transform as
\begin{align}
    c_i^\dagger c_i &=S_i^+ S_i^-  \\
    c_{i+1}^\dagger c_i &= S_{i+1}^+ S_i^-, \\
    c_{i+1}^\dagger c_i^\dagger &= -S_{i+1}^+ S_i^+.
\end{align}
After this transformation, the fermionic states \(\ket{0}\) and \(\ket{1}\), representing the vacant or occupied site, correspond to spin-down $\ket{\downarrow}$ and spin-up \(\ket{\uparrow}\), respectively. This establishes a direct correspondence between the fermionic and spin bases. Finally, the Hamiltonian in equation~\eqref{ham} is transformed into spin Hamiltonian
\begin{equation}
    \hat{H} = -\frac{\mu N}{2} - \mu \sum_{i=1}^{N} S_i^z 
    + 2 \sum_{i=1}^{N-1} \left[(t + \Delta) S_{i+1}^y S_i^y 
    + (t - \Delta) S_{i+1}^x S_i^x\right], \label{JW_H}
\end{equation}
expressed in terms of the spin operators $S_{i}^{x}$, $S_{i}^{y}$, and $S_{i}^{z}$. By comparison with the XY model
\begin{equation}
    \hat{H} = - h \sum_{i=1}^{N} S_i^z 
    +  \sum_{i=1}^{N-1} [J_y S_{i+1}^y S_i^y 
    + J_x S_{i+1}^x S_i^x], \label{XY_H}
\end{equation}
we see that the Kitaev Hamiltonian in spin representation is essentially an XY model, with $\mu$ acting as the external magnetic field along the $z$-direction, $J_x$ as the $t-\Delta$ term, and $J_y$ as the $t+\Delta$ term.
\begin{figure}[]
\centering
  \includegraphics[width=0.7\linewidth]{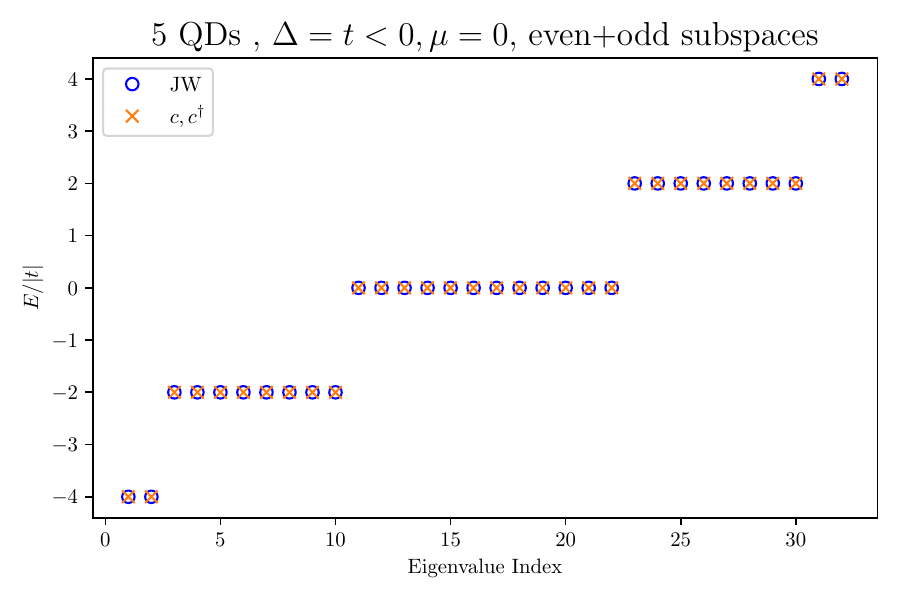}
  \caption{Energy spectra matching of the fermionic and spin Hamiltonian in the analytical solution case. The even (odd) eigenvalue indices correspond to the even (odd) subspace eigenvalues. }
  \label{JW}
\end{figure}
In Figure \ref{JW} we show results using ED of the fermionic Kitaev Hamiltonian and spin Kitaev Hamiltonian obtained through JW transformation. This figure shows that the spectra are identical and we can proceed to implementing VQE on a quantum computer.

\subsection{Quantum Circuit and Variational Wave Function}
We utilize four ansätze (quantum circuits) throughout the variational calculation, namely Even, Odd, Efficient SU(2), and Hamiltonain variational ansatz (HVA) \cite{wiersema2020exploring, wecker2015progress}. Inspired by the fact that in the TP, the ground state of even (odd) subspace has equal contribution of all even (odd) configurations, we construct the even (Odd) ansatz based on this feature, i.e. we consider all configurations in the subspace, and assign variational parameters to them.  
\begin{figure}[h]
\begin{center}
\begin{quantikz}
\lstick{$\ket{0}$} & \gate{R_x(\theta_1)} & \ctrl{1}  & \qw       & \qw       & \targ{}  & \gate{R_z(\theta_4)} & \qw \\
\lstick{$\ket{0}$} & \gate{R_x(\theta_2)} & \targ{}   & \ctrl{1}  & \targ{}   & \ctrl{-1}  & \qw & \qw \\
\lstick{$\ket{0}$} & \gate{R_x(\theta_3)} & \qw       & \targ{}   & \ctrl{-1}  & \qw       & \qw & \qw
\end{quantikz}
\caption{The quantum circuit for the \textbf{even} ansatz with 3 qubits. The gates are applied as: a set of \( R_x \) gates applied to each qubit. Then, CNOT gates are applied in a forward direction between consecutive qubits. After that, reverse CNOT gates are applied in reverse order. Finally, a single \( R_z \) gate is applied only to the first qubit (qubit 0).}\label{even_ansatz_circuit}
\end{center}
\end{figure}
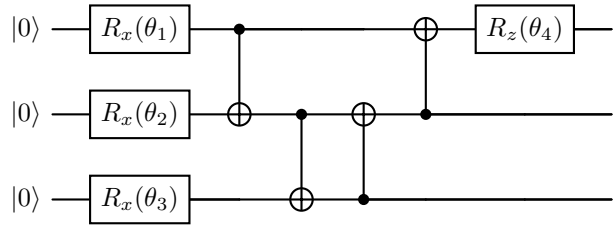
The Efficient SU(2) consists of layers of single qubit operations spanned by the SU(2) group and CX entanglements. HVA is directly built from the Hamiltonian (see Appendix \ref{app1}). As an illustration, we look at the even ansatz for 3 qubits in Figure \ref{even_ansatz_circuit}. This ansatz essentially creates all the even configurations using the $R_y$ and CNOT operations, and assigns variational parameters to each configuration.
\begin{equation}
    \ket{\psi({\boldsymbol{\theta}})}_{\text{even}} = R_{Z}^{(0)}(\theta_{N+1}) \prod_{j=1}^{N-1}\text{CX}^{(j+1, j)} \prod_{j=1}^{N-1} \text{CX}^{(N-j, j-N+1)} \prod_{j=1}^{N} R_{X}^{(j)}(\theta_{j}) \ket{\psi_0}
\end{equation}

\section{Results of VQE}
\subsection{Comparison of Quantum Circuits for Variational Ansätze}
In order to make a transparent comparison between the performance of different ansätze, we examine the loss function behaviour, which in this case is the ground state energy.

\begin{figure}[H]
    \centering
    \begin{subfigure}[t]{0.45\textwidth}
        \centering
        \begin{overpic}[width=\textwidth]{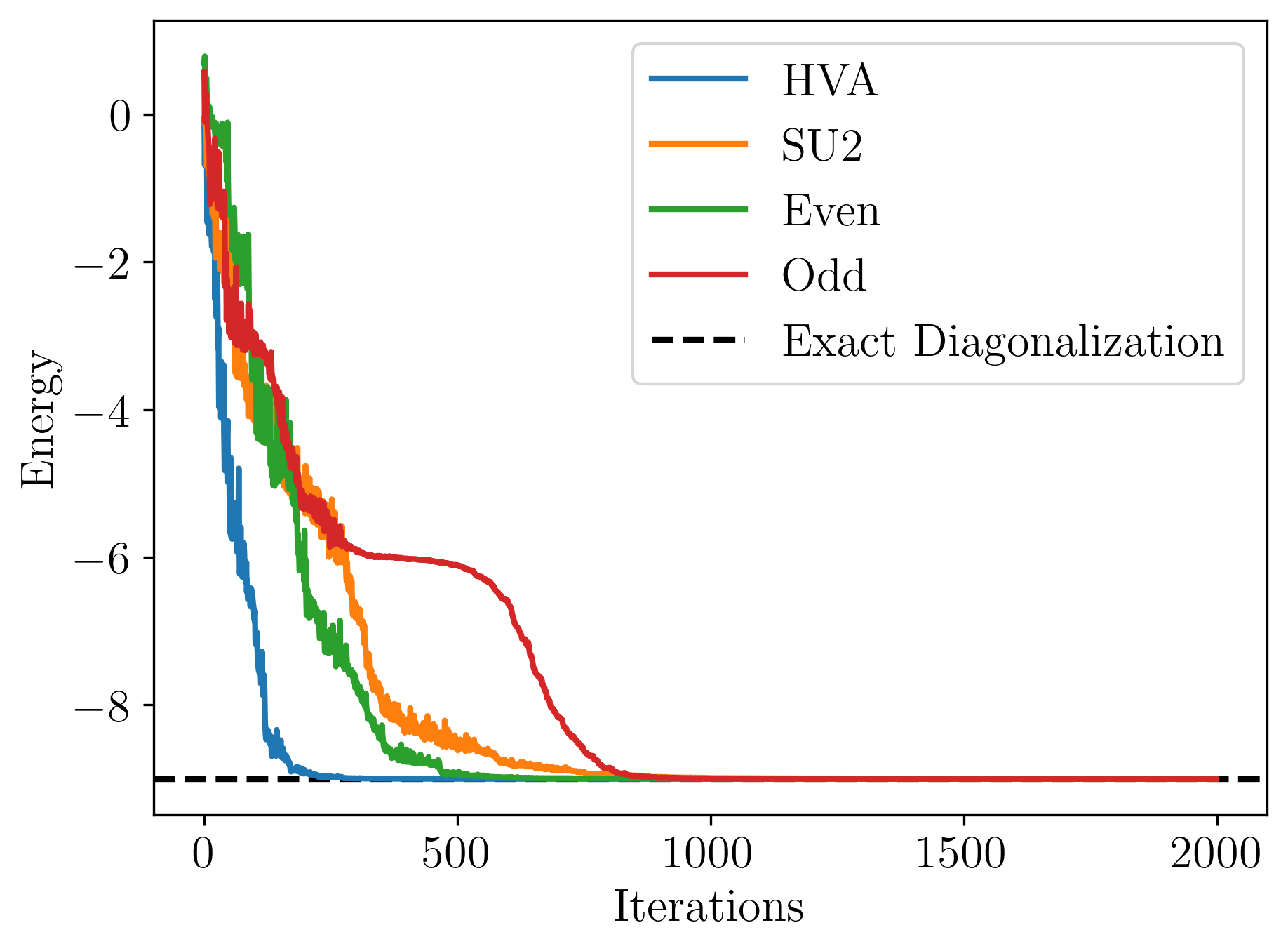}
            \put(90, 20){(a)} % Adjust coordinates (x, y) as needed
        \end{overpic}
        \phantomcaption
        \label{loss_anat}
    \end{subfigure}
    % \hfill
    \begin{subfigure}[t]{0.45\textwidth}
        \centering
        \begin{overpic}[width=\textwidth]{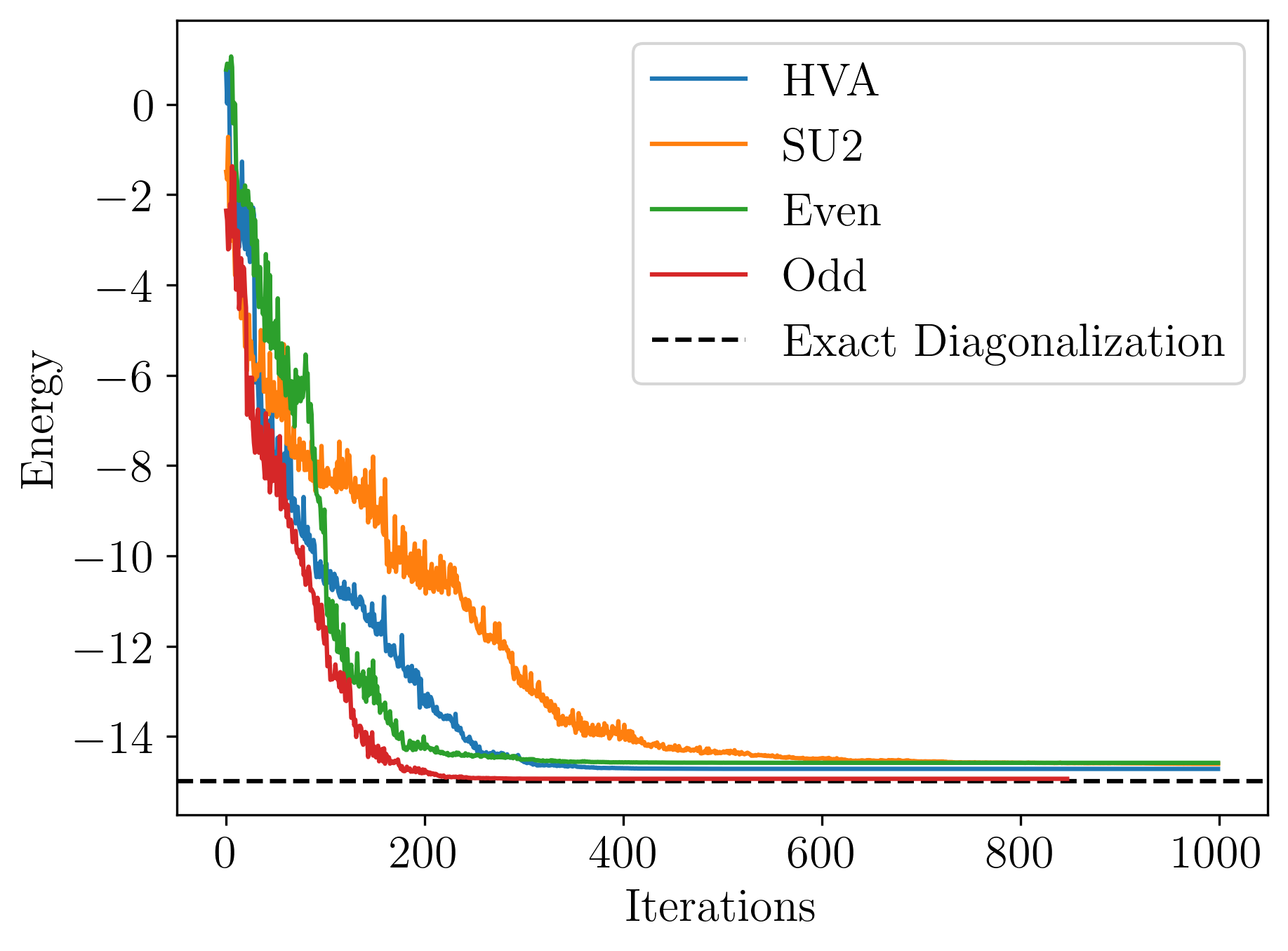}
            \put(90, 20){(b)} % Adjust coordinates (x, y) as needed
        \end{overpic}
        \phantomcaption
        \label{loss_no_anat}
    \end{subfigure}
    \caption{Ground state energy as a function of optimization steps for four ansätze: Even, Odd, SU2, and HVA, for a chain of 10 QDs. The comparison highlights the convergence pattern and efficiency of each ansatz in minimizing the ground state energy. (a) corresponds to the analytical case with $\Delta = t$ and $\mu = 0$, while (b) represents a case outside the analytical regime but still within the topological phase, with $\Delta = 2t$ and $\mu = -0.2t$. The optimal number of parameters is used in both cases.}
    \label{loss_two}
\end{figure}

As shown in Figure \ref{loss_two}, although HVA is the fastest to reach convergence, the even, SU2, and odd ansätze have smaller error, especially in the non-analytical solution case. Each ansatz stops at a particular optimization step once the classical optimizer decides it has converged. For each ansatz, the best number of parameters is used from studying the relative error behaviour for various number of parameters, discussed in Appendix \ref{app2}.

In Figure \ref{loss_anat}, the performance of each ansatz is shown in the analytical solution regime. The Ground State Energy for even, odd, and SU2 ansätze is closer to the exact ground state energy, compared to HVA.

In Figure \ref{loss_no_anat} although there is no analytical solution for the choice of parameters, the system still lies in the TP. The exact ground state energy is obtained using numerical ED. The performance of each ansatz is displayed for their best number of parameters in Apendix \ref{app2}. Although the optimization process is more challenging in this case, the ground state energy for even, odd, and SU2 ansätze indicate reliable performance, while HVA stops in the early optimization steps, with a significant error.

This reveals a clear distinction in the performance of the different ansätze when comparing the relative error. For the even, odd, and SU2 ansatzes, convergence to the actual ground state energy is achieved with an insignificant relative error. In contrast, despite the fast convergence, the HVA ansatz shows a higher relative error, specifically in the non-analytical case. 

\subsection{Error Sensitivity within the Topological Phase}
Having demonstrated the convergence of the variational functions, we now explore the sensitivity of the approximated ground state energy as the parameters scale up in the TP. To do so, we analyze the relative error for each ansatz while we tune the parameters within the TP in the parameter space.
\begin{figure}[H]
\centering
  \includegraphics[width=0.7\linewidth]{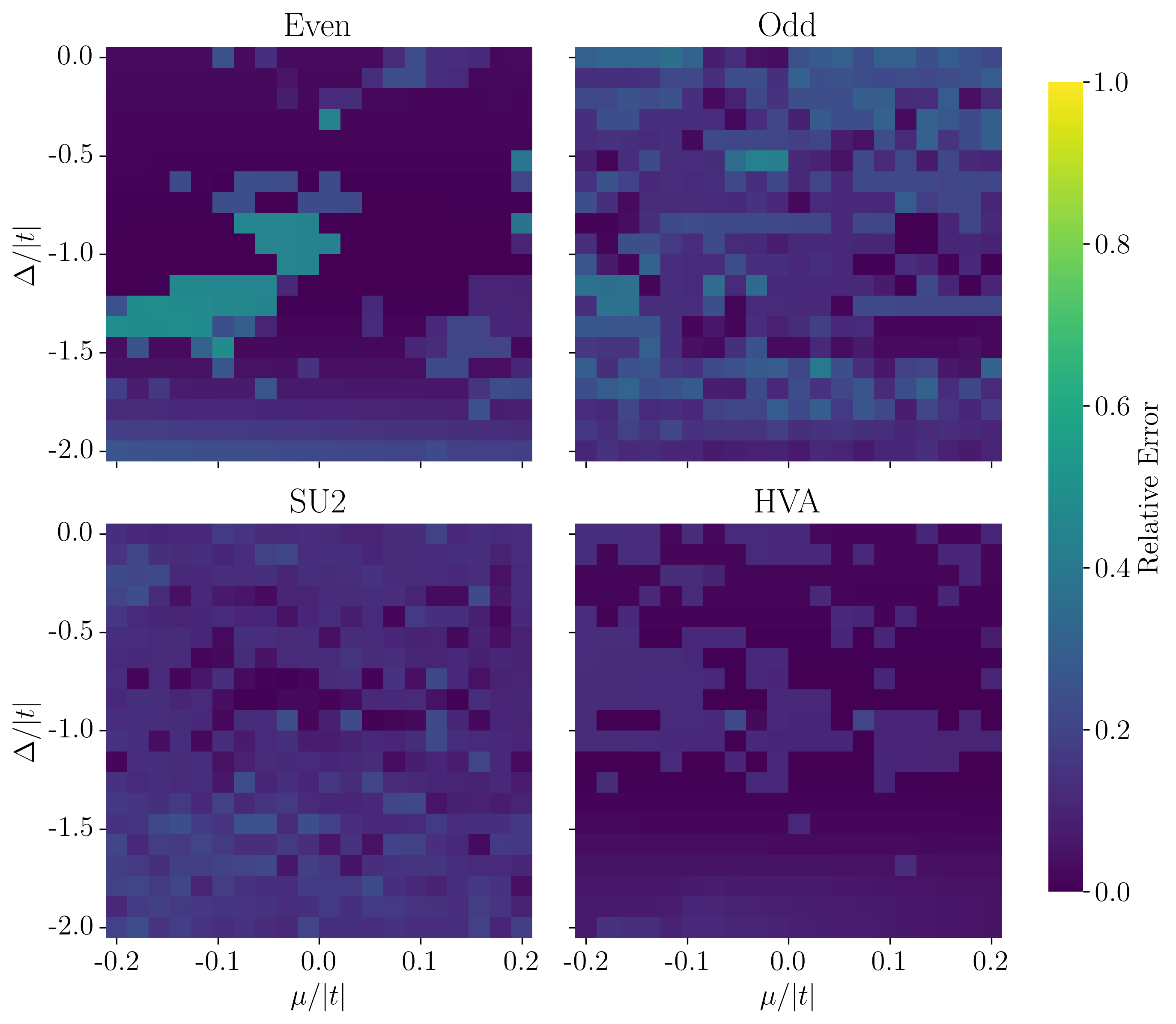}
  \caption{  Evolution of the relative error in the parameter space for various ansätze, for a chain of 10 QDs. The choice of the parameters $\Delta/|t|$ and $\mu/|t|$ is such that the system remains in the TP. The dark regions indicate areas of lower relative error, with the analytical solution located at the center of the plots.}
   \label{heat}
\end{figure}
Figure \ref{heat} visualizes the performance of each variational ansatz by displaying the relative error as a function of $\Delta/|t|$ and $\mu/|t|$. The special case where the analytical solution case is at the centre of each figure. Except for the HVA case, as the parameters deviate from this case, the relative error increases. Each ansatz exhibits a unique error profile, highlighting differences in how well they adapt to variations in the Hamiltonian. However, the even, odd, and SU2 ansätze outperform HVA. This disparity of performance  highlights the suitability of the even, odd, and SU2 ansätze for accurately modeling the system compared to more complex alternatives like the HVA.

\section{Conclusion}
In summary, we presented steps toward designing Majorana Quasiparticles in InAsP Quantum Dots in InP Nanowires on p-wave superconductor with  QNANO and Variational Quantum Eigenvalue Solver on a quantum computer. We used VQE  approach to estimate the ground state energy of the seminconducting QD nanowire in the proximity of a p-wave superconductor, in the presence of applied external magnetic field, described by the Kitaev chain. We used QNANO to obtain single particle spectra and showed that by constructing the optimal quantum circuit for variational wave function from the analytical solution of the system, we can determine the odd and even ground state energy and hence determine the topological phase of the Kitaev chain. Moreover, the topological features of the wave function remain preserved, thus enabling the system to host MZMs. Future work will focus on simulating spinfull fermions in semiconducting InAs quantum dots in InP nanowires with Rashba spin-orbit coupling on s-type superconductors. 

\medskip
\textbf{Supporting Information} \par %Please delete the Suppporting Information statement if it is not applicable. Please supply Supporting Information in another file. Supporting information should not be provided in .tex format
Supporting Information is available from the Wiley Online Library or from the author.

% Acknowledgements
\medskip
\textbf{Acknowledgements} \par %delete if not applicable))
P.H., M.M, DM, IA, IC, AWR were supported by the AQC-004 project of the
Applied Quantum Computing Program at the National Research Council
of Canada,  NSERC Discovery Grant No. RGPIN 2019-05714, NSERC Alliance Quantum
Grant No. ALLRP/578466-2022, the QSP-078 project of the
Quantum Sensors Program at the National Research Council
of Canada, University of Ottawa Research Chair in Quantum
Theory of Materials, Nanostructures, and Devices, and Digital
Research Alliance Canada with computing resources.

\textbf{Conflicts of Interest}\par
The authors declare no conflict of interest.

%Appendix
\appendix
\section{Ansatz Construction} \label{app1}
\begin{figure}[h]
    \begin{center}
        \begin{quantikz}
            \lstick{$\ket{0}$} & \gate{R_y(\theta_1)} & \gate{R_z(\theta_2)} & \ctrl{1} & \qw & \qw & \gate{R_y(\theta_7)} & \gate{R_z(\theta_8)} \\
            \lstick{$\ket{0}$} & \gate{R_y(\theta_3)} & \gate{R_z(\theta_4)} & \targ{} & \ctrl{1} & \qw & \gate{R_y(\theta_9)} & \gate{R_z(\theta_{10})} \\
            \lstick{$\ket{0}$} & \gate{R_y(\theta_5)} & \gate{R_z(\theta_6)} & \qw & \targ{} & \qw & \gate{R_y(\theta_{11})} & \gate{R_z(\theta_{12})}
        \end{quantikz}
        % Add the underbrace below the first two layers of gates
        \begin{tikzpicture}[overlay, remember picture]
            % Position the underbrace and text
            \draw[thick, decorate, decoration={brace, amplitude=5pt, mirror}] 
                (-9.6,-1.5) -- (-6.1,-1.5) node[midway, below=5pt] {\footnotesize don't repeat};
        \end{tikzpicture}
    \end{center}
    \caption{Quantum circuit of the SU2 ansatz with 3 qubits and 1 layer of linear entanglement.}
    \label{SU2_circuit}
\end{figure}
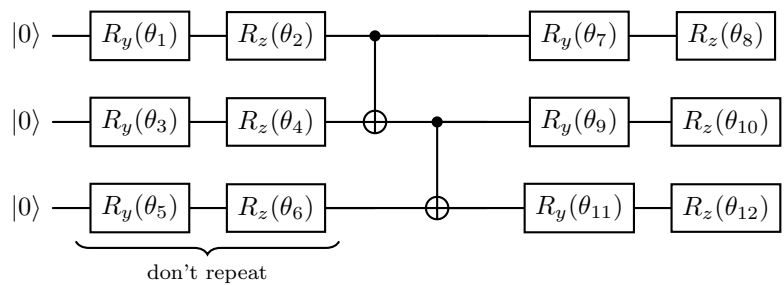

As mentioned before, we use four different ansätze, namely even, odd, Efficient SU(2), and HVA. The even ansatz is already described in Figure \ref{even_ansatz_circuit}. The odd ansatz is a variational function that contains all the odd subspace configurations (Figure \ref{odd_ansatz_circuit})
\begin{equation}
    \ket{\psi({\boldsymbol{\theta}})}_{\text{odd}} = \prod_{j=1}^{N} R_{Z}^{(j)}(\theta_{N+j}) \prod_{j=1}^{N-1}\text{CX}^{(j+1, j)} \prod_{j=1}^{N-1} \text{CX}^{(N-j, j-N+1)} \prod_{j=1}^{N} R_{X}^{(j)}(\theta_{j}) \ket{\psi_0}
\end{equation}
\begin{figure}[h]
\begin{center}
\begin{quantikz}
\lstick{$\ket{0}$} & \gate{R_y(\theta_1)} & \ctrl{1}  & \qw       & \qw       & \targ{}  & \gate{R_z(\theta_4)} & \qw \\
\lstick{$\ket{0}$} & \gate{R_y(\theta_2)} & \targ{}   & \ctrl{1}  & \targ{}   & \ctrl{-1}  & \gate{R_z(\theta_5)} & \qw \\
\lstick{$\ket{0}$} & \gate{R_y(\theta_3)} & \qw       & \targ{}   & \ctrl{-1}  & \qw       & \gate{R_z(\theta_6)} & \qw
\end{quantikz}
\caption{The quantum circuit for the \textbf{odd} ansatz with 3 qubits. The gates are applied as: a set of \( R_y \) gates applied to each qubit. Then, CNOT gates are applied in a forward direction between consecutive qubits. After that, revers CNOT gates applied in the reverse order. Finally, a set of \( R_z \) gate is applied to each qubit.
}
\label{odd_ansatz_circuit}
\end{center}
\end{figure}
The Efficient SU(2) (Figure \ref{SU2_circuit}), namely from the special unitary group, consists of layers of single-qubit rotations ($R_y$ and $R_z$ gates) and entangling gates (CNOT) arranged to capture correlations between qubits. By applying these operations across multiple layers, the ansatz can approximate the ground state of complex quantum systems with fewer resources than more traditional methods.

\begin{equation}
    \ket{\psi({\theta})}_{\text{SU2}} = \prod_{j=1}^{N} R_{Y}^{(j)}(\theta_{2j-1})R_{Z}^{(j)}(\theta_{2j}) \prod_{j=1}^{N-1}\text{CX}^{(j+1, j)} \prod_{j=1}^{N} R_{Y}^{(j)}(\theta_{2N+2j-1})R_{Z}^{(j)}(\theta_{2N+2j}) \ket{\psi_0}
\end{equation}

The last ansatz we consider is the HVA. To construct it, we decompose the Kitaev Hamiltonian, express in Equation (\ref{JW_H}), in to a sum of non-communing terms
\begin{equation}
        \hat{H} = \frac{t-|\Delta|}{2} \sum_{j=1}^{N-1} \left( \sigma_{j+1}^{x} \sigma_{j}^{x} \right) + \frac{t + |\Delta|}{2} \sum_{j=1}^{N-1} \left( \sigma_{j+1}^{y} \sigma_{j}^{y} \right) - \frac{\mu}{2} \sum_{j=1}^{N} \sigma_{j}^{z} - \frac{\mu N}{2}.
\end{equation}
For a decompose Hamiltonian \(\hat{H} = \sum_{i=1}^{q} c_{j}\hat{H}^{(j)}\), where \(\{c_j\}\) are the real coefficient, each term in the Hamiltonian corresponds to a unitary variational operator \(\hat{U}_j = \exp(-i\theta_j \hat{H}^{(j)})\), associated with a variatinoal parameters \(\theta_j\), that contributes to the overall HVA circuit. Then, the Kitaev HVA is constructed as
\begin{equation}
    \hat{U}(\boldsymbol{\theta}) = \prod_{k=1}^{R}\left[ \prod_{j=1}^{N-1}\exp\left\{ -i \theta_{j}^{x} \left( \sigma_{j+1}^{x} \sigma_{j}^{x} \right) \right\} 
    \prod_{j=1}^{N-1}\exp\left\{ -i \theta_{j}^{y} \left( \sigma_{j+1}^{y} \sigma_{j}^{y} \right) \right\}
    \prod_{j=1}^{N}\exp\left\{ -i \theta_{j}^{z} \sigma_{j}^{z} \right\} \right]
\end{equation}
Another step it to apply a second-order Trotter-Suzuki decomposition to implement each exponential operator \(\hat{U}_j\). This method approximates the time evolution operator by applying each term in sequence, alternating the order of application to reduce errors, from what is  already established \cite{wecker2015progress}.

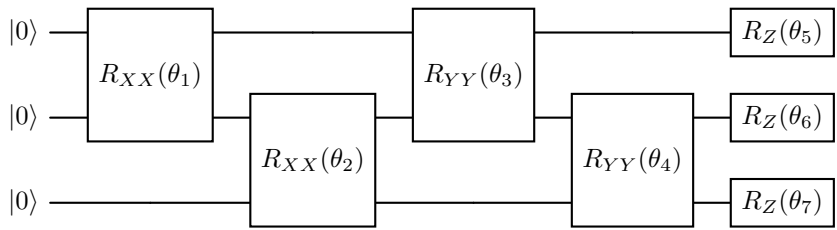
\begin{figure}[h]
\begin{center}
\begin{quantikz}
\lstick{$\ket{0}$} & \gate[2]{R_{XX}(\theta_1)} & & \gate[2]{R_{YY}(\theta_3)} & & \gate{R_Z(\theta_5)} \\
\lstick{$\ket{0}$} &  & \gate[2]{R_{XX}(\theta_2)} &  & \gate[2]{R_{YY}(\theta_4)} & \gate{R_Z(\theta_6)} \\
\lstick{$\ket{0}$} &  &  &  &  & \gate{R_Z(\theta_7)}
\end{quantikz}
\caption{The quantum circuit for the HVA with 3 qubits}\label{even_ansatz_circuit}
\end{center}
\end{figure}

\section{Ansatz Optimization} \label{app2}
To improve the performance of each ansatz, we optimize them based on the number of layers and how we can minimize the relative error. To do so, we plot the relative error vs. the number of layers (parameters) of each ansatz. For a given number of layers, we take the minimum energy of 10 initial runs. The entire optimization is done using the \textbf{COBYLA} optimizer available in the \textbf{Qiskit} package, and 30000 shots.

\begin{figure}[H]
\centering
  \includegraphics[width=0.6\linewidth]{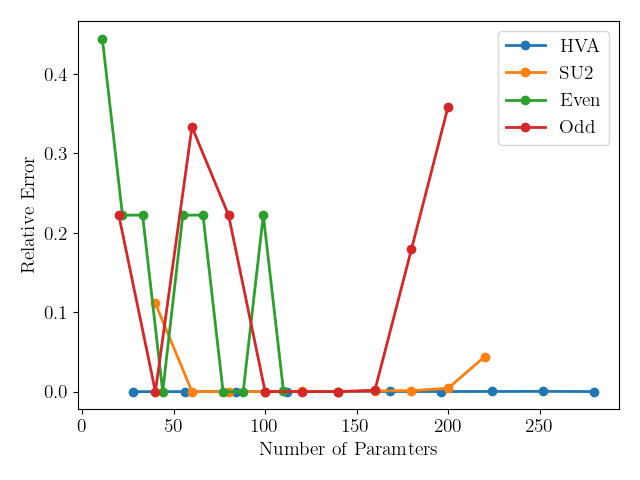}
  \caption{Relative error as a function of the number of parameters for different ansätze in the analytical solution regime, $\Delta = t$ and $\mu=0$, for a chain of 10 QDs. The maximum number of layers is 10.}
  \label{params}
\end{figure}

In Figure \ref{params}, each ansatz is evaluated across varying numbers of layers, with the number of layers ranging from 1 to 10. The relative error is plotted on the y-axis,
\begin{equation}
    \text{Error} = \left| 1 - \frac{E_{\text{VQE}}}{E_{\text{ED}}} \right|,
\end{equation}
while the number of parameters on the x-axis, allowing for a direct comparison of how the error changes with increasing the number of layers for each ansatz. The number of parameters per layer for the even, odd, SU2, and HVA ansätze are 11, 20, 40, and 28, respectively.

% References
\medskip

% Use the following code if you wish to generate your bibliography with BibTeX;
% replace the string "MSP-template" below with the name(s) of
% the BibTeX data base(s) you want to use.
% The resulting bibliography-output (the content of the .bbl file)
% must be pasted back into this file before submission.
% Please also include your BibTeX data base file(s) in your submission
% so that we can re-run BibTeX if necessary.
%
\bibliographystyle{MSP}
%\bibliography{MSP-template}


\bibliography{ref}
%\bibliographystyle{chicago}

% Figures/tables and captions
% Permission statements are required for all figures reproduced or adapted from previously published articles/sources. Please also ensure that all necessary permissions to reproduce images have been received
% Please remove these statements for original figures

% Please provide Biographies and photos for Essays, Feature Articles, Progress Reports, Reviews, and Perspectives for those authors who should be highlighted  
% These should be at most 100 words long
% For other article types this section can be removed
% Photographs should be 40mm broad and 50 mm high

% Table of contents entry should be 50 - 60 words long
% Image should be 55 mm broad and 50 mm high or 110 mm broad and 20 mm high

\end{document}